\documentclass[CEJP,PDF]{cej} 
\usepackage{layout}
\usepackage{amsmath}
\usepackage{textcomp}
\usepackage{hyperref}

\usepackage{graphicx}
\usepackage{psfrag,graphics}
\usepackage{amsmath}
\usepackage{subfigure}
\usepackage{lineno}

\usepackage{mathtools}
\usepackage{setspace}

\title{The Horn, Kink and Step, Dale: \\from few GeV to few TeV}

\articletype{Research Article}

\author{Anar Rustamov\inst{1}\email{a.rustamov@cern.ch}}

\institute{
     \inst{1}Goethe-Universit\"at Frankfurt\\
    Max-von-Laue-Str. 1, 60438 Frankfurt am Main, Germany
             }

\abstract{Rich experimental data have been collected in heavy-ion collisions at high energies to study the properties of strongly interacting matter. As the theory of strong interactions, QCD, predicts asymptotic freedom, the created matter at sufficiently high temperature and density will be dominated by a state of quasi-free quarks and gluons referred to as the Quark-Qluon Plasma (QGP). Experimental signals for the onset of the QGP creation (the onset of the deconfinement) have been predicted within the statistical model for the early stage of nucleus-nucleus collisions. In this model the existence of two different phases is assumed: confined mater and the QGP, as well as a first order phase transition between them.  Until recently, these predictions were confirmed only by the NA49 experiment at the CERN SPS. In this report recent results from STAR at RHIC/BNL and from ALICE at LHC/CERN, related to the onset of deconfinement, will be compared to published results from NA49.}

\keywords{heavy-ion, quark gluon plasma, phase transitions, onset of deconfinement}
\pacs{25.75.-q, 25.75.Nq}

\begin{document}
\maketitle


The systematic study of hadron production in central Pb+Pb collisions performed by the NA49 collaboration~\cite{na49_1} revealed structures in energy dependence of the following observables:
\begin{enumerate}
\item The Horn: A sharp maximum in  $<K^{+}>/<\pi^{+}>$ ratio,
\item The Kink:  Sudden change in the number of pions per participant,  
\item The Step: A plateau in the energy dependence of the inverse slope parameter.
\end{enumerate}
These structures have been predicted within the statistical model for the early stage (SMES)~\cite{marek_1, marek_2} of nucleus-nucleus collisions as a hint for the onset of the QGP creation. 
The SMES, based on the seminal paper of Fermi~\cite{fermi_1}, assumes that depending on collision energy the produced system appears in one of two states: a thermalized hadron gas (confined phase) or a thermalized QGP state (deconfined phase). Adopting the Fermi-Landau initial conditions the energy density of the system is related to the Mandelstam variable, s\footnote{Hereafter the notation $s_{NN}$ is used, defined for the system of two nucleons.}. At the critical temperature a first order phase transition between these two phases is assumed. Furthermore, during expansion of the system the entropy and strangeness is assumed to be constant.

\begin{figure}[htb]
\includegraphics[width=7.5cm]{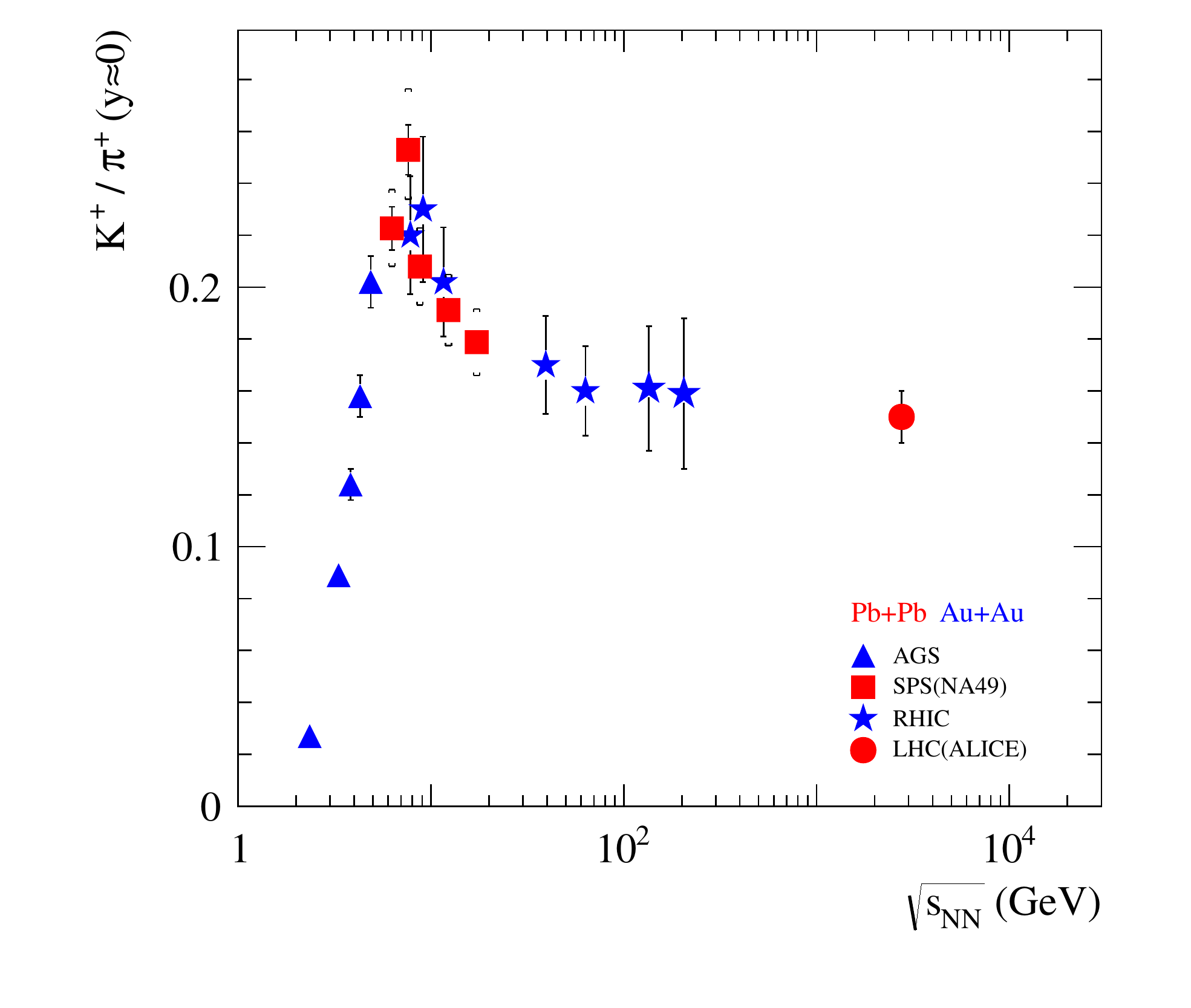}
\includegraphics[width=7.5cm]{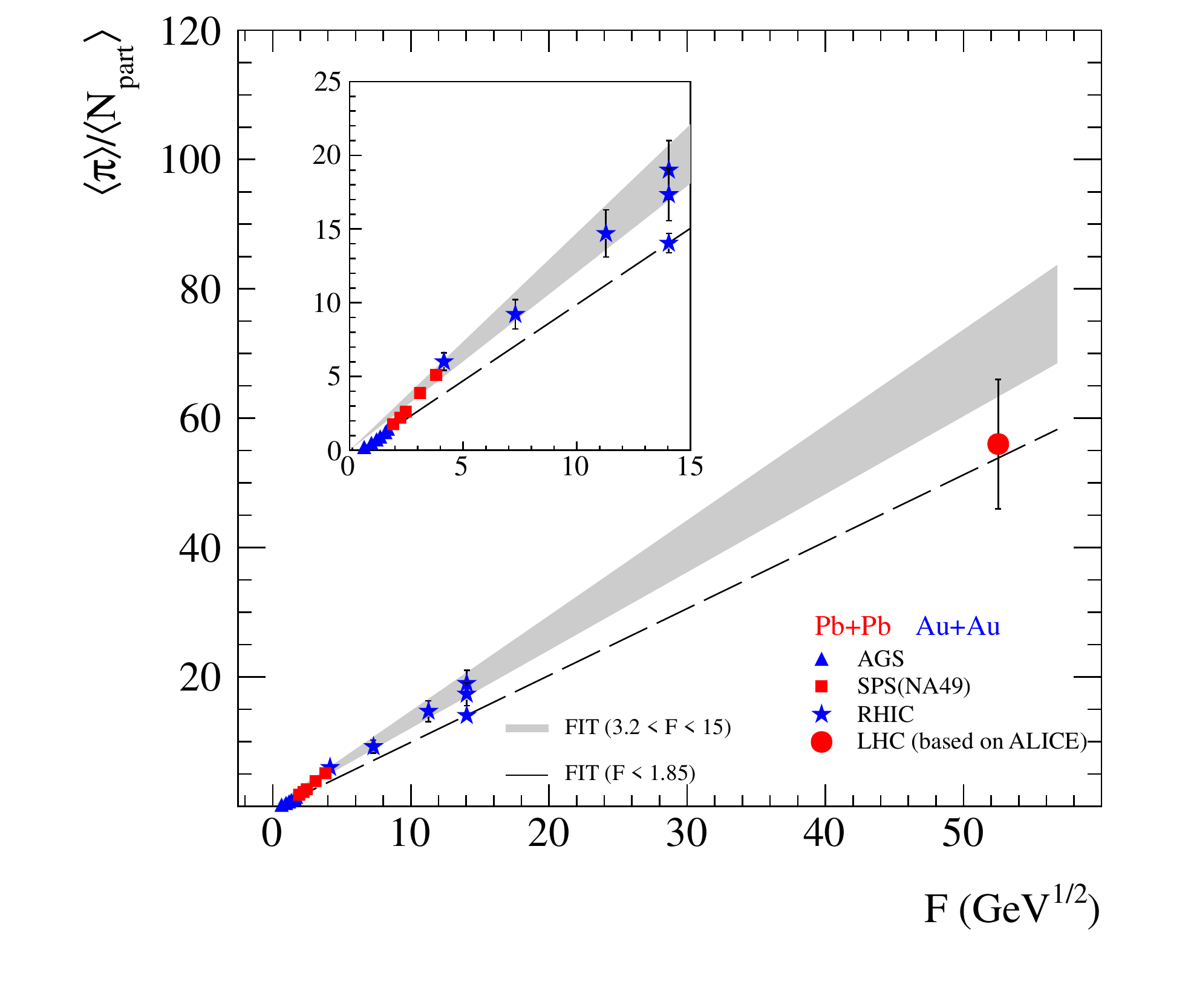}
\caption{(Color online) Left panel: $\left<K^{+}\right>/\left<\pi^{+}\right>$ ratio measured in central Pb+Pb (red symbols) and Au+Au (blue symbols) collisions as a function of Mandelstam variable, $s_{NN}$. An evident peak observed by the NA49 collaboration is confirmed by the STAR experiment. The ALICE point is consistent with the established trend. Right panel: Pion multiplicity normalized to the number of participant nucleons as a function of the Fermi variable, F$\approx s_{NN}^{1/4}$.}
\label{fig_horn}
\end{figure}

Recently the STAR collaboration reported on new measurements in central Au+Au collisions~\cite{star_1}. Moreover, ALICE has presented $<K>/<\pi>$ ratio in Pb+Pb collisions at 2.76 TeV~\cite{ALICE_1}. The current status of the "Horn plot" is presented in the left panel of Fig.~\ref{fig_horn}.
New measurements of STAR are in a reasonable agreement with the NA49 findings. Furthermore, the ALICE point serves as an important confirmation of the established and predicted trend.
Even though the statistical hadron gas model~\cite{pbm_1} also exhibits an enhancement close to the peak structure, its width is by far too large compared to the experimental observations. It should, however, be noted that recently by incorporating higher-lying Hagedorn states and the scalar $\sigma$ meson in their model, the authors of~\cite{pbm_2}  report on a significantly improved fit to the observed structure. 

The right panel of the Fig.~\ref{fig_horn} demonstrates the current status of the "Kink plot". An anticipated change of the slope is better seen in the inset of the plot. The number of charged particles extrapolated to full phase space, $N_{ch.}^{total}=17100\pm1000$, measured in central Pb+Pb collisions at $\sqrt{s_{NN}}=$2.76 TeV was recently published by the ALICE collaboration~\cite{ALICE_2}. Together with the values of the $<K^{-}>/<\pi^{-}>$ and $<\bar{p}>/<\pi^{-}>$ ratios, reported in~\cite{ALICE_1}, the pion multiplicity in full phase space is  estimated using the following assumptions:

\begin{equation}
\frac{K^{+}+K^{-}}{\pi^{+}+\pi^{-}}=\frac{K^{-}}{\pi^{-}}; \ \ \ \ \ \frac{p+\bar{p}}{\pi^{+}+\pi^{-}}=\frac{\bar{p}}{\pi^{-}},
\end{equation}


\begin{equation}
\left<N_{ch.}^{total}\right>=\left<\pi^{+}+\pi^{-}\right>+\left<K^{+}+K^{-}\right>+\left<p+\bar{p}\right>+\frac{N_{part}}{2},
\end{equation}
where $N_{part}$ is a number of participants.

Using Eq. (2) and the isospin symmetry of the colliding system, the total number of the pions normalized to the number of participating nucleons is obtained:

\begin{equation}
\frac{\left<N_{\pi}\right>}{N_{part}}=\frac{1.5\left<\pi^{+}+\pi^{-}\right>}{384}=55.26\pm9.26.
\end{equation}

The quoted error is the result of statistical error propagation with the systematic one added in quadrature. The latter is taken to be 15$\%$  due to approximations described in Eqs. (1) and  (2). 
This point is presented in Fig.~\ref{fig_horn}  by the full red circle. It is remarkable that within error bars the point is close to the predictions marked by the shaded area. 



\begin{figure}[htb]
\includegraphics[width=7.5cm]{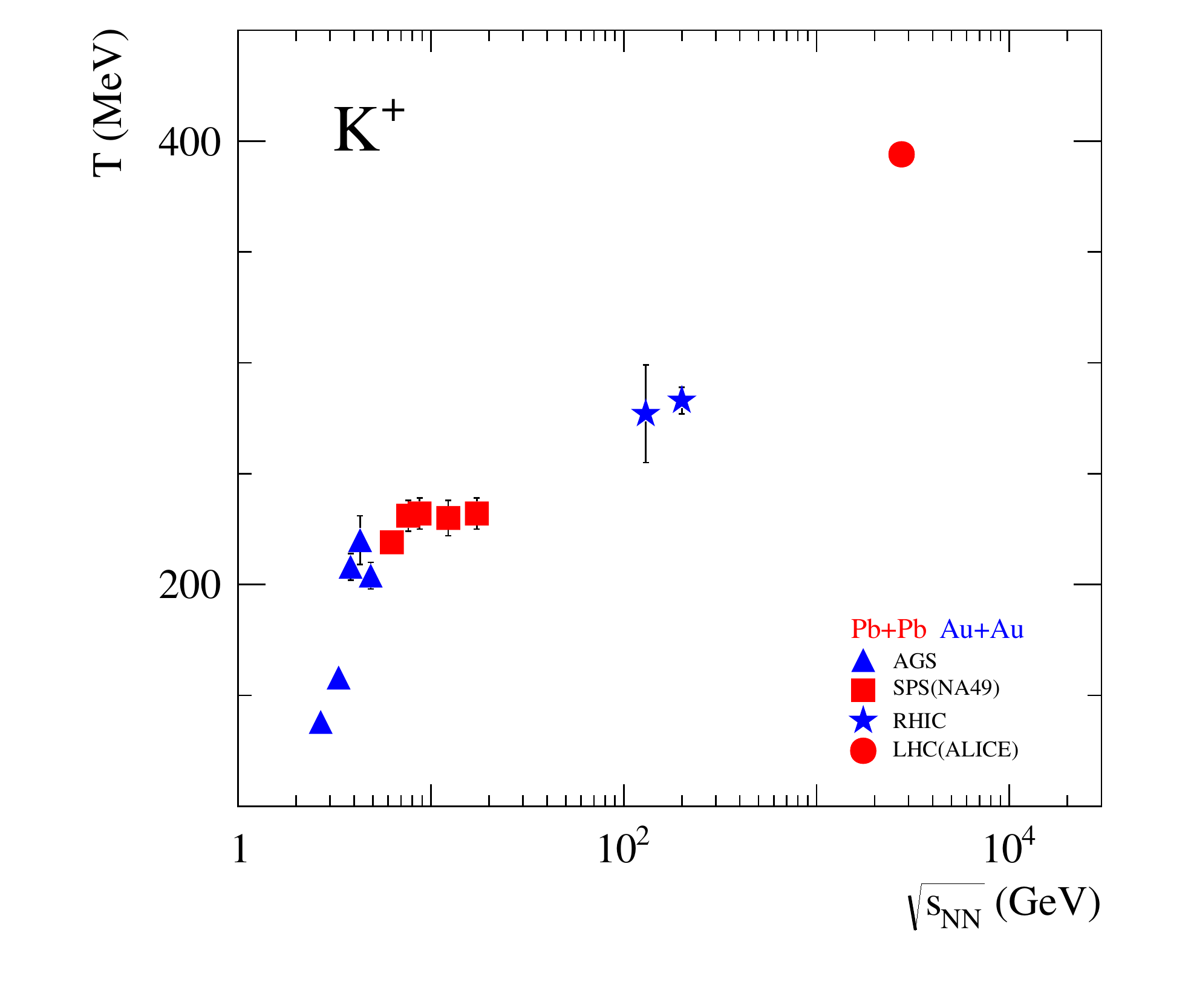}
\includegraphics[width=7.5cm]{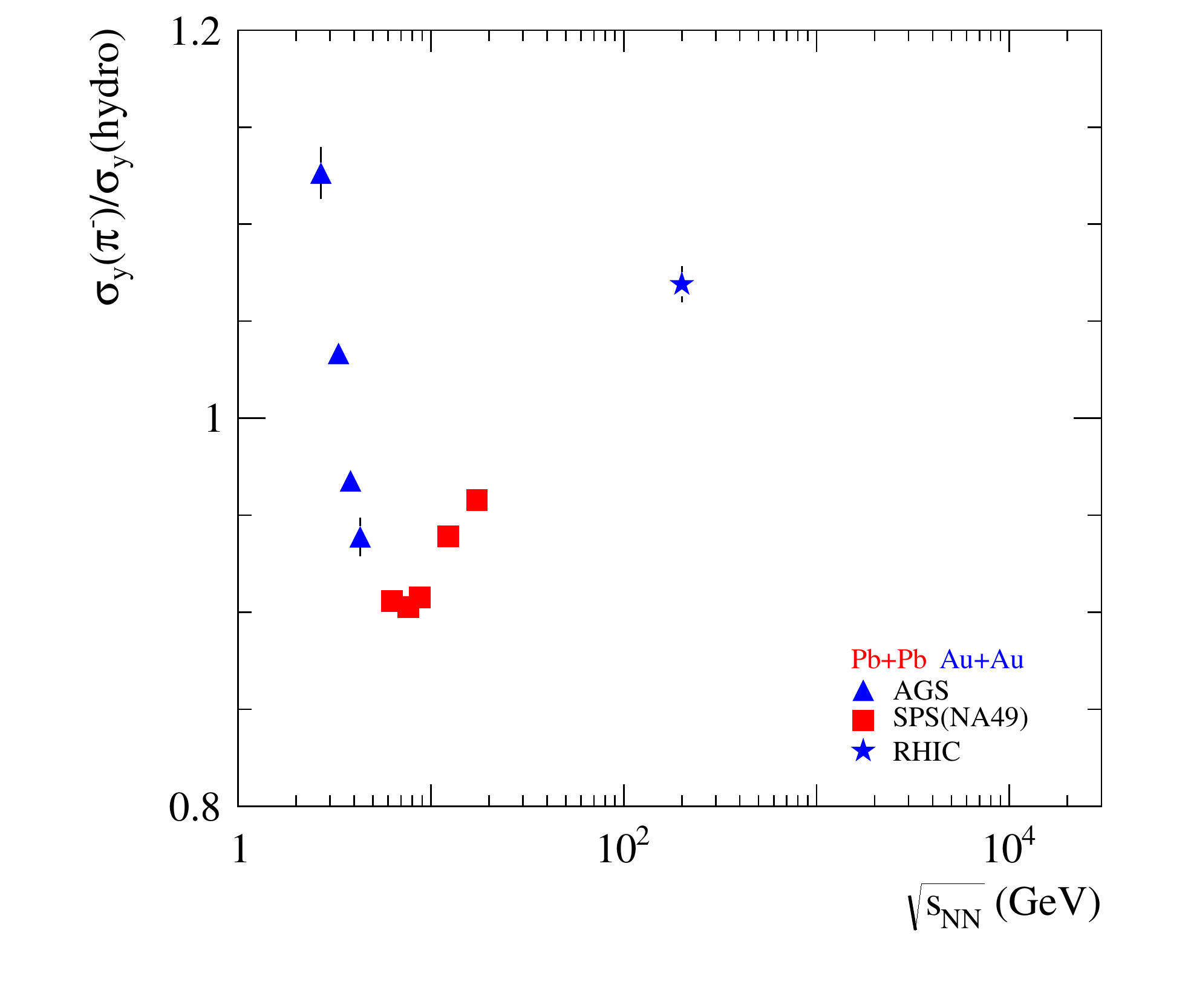}
\caption{(Color online) Left panel: Inverse slope parameter of the $m_{t}$ distribution as a function of Mandelstam variable, $s_{NN}$. With the inclusion of the ALICE point the plateau behavior is established. Right panel: The experimentally measured widths of the pion rapidity distribution normalized to the calculations within Landau's hydrodynamical model. The step and dale structures indicate softening of the EoS in the SPS energy range as it was  predicted for the onset of deconfinement.}
\label{fig_step}
\end{figure}

Experimental transverse mass spectra, $m_{t}=\left(p_{t}^2+m^{2}\right)^{1/2}$, are usually parametrized with an exponential function:
\begin{equation}
\frac{d^{2}N}{m_{t}dm_{t}dy}=Ce^{-\frac{m_{t}}{T}}.
\end{equation}
In hydrodynamic models the inverse slope parameter, T, is determined by the temperature, pressure and freeze-out conditions.
The Step, i.e. a plateau in the energy dependence of the inverse slope parameter of charged kaons is a natural consequence of the assumption of the first order phase transition (mixed phase) in SMES.  As the temperature as well as the pressure in the mixed phase are constant the observed plateau is attributed to the reduction of the transverse flow, i.e to softening of the Equation of State (EoS). An indication of such a plateau was again observed by the NA49 Collaboration~\cite{na49_1}. In order to obtain the corresponding LHC point, the transverse momentum distribution of the charged kaons measured by ALICE~\cite{ALICE_1} was transformed first into transverse mass distribution and then fitted by exponential function defined in Eq. (4). The extracted inverse slope parameter is plotted in Fig.~\ref{fig_step}. The plateau behavior in the energy dependence of the inverse slope parameter is confirmed by this plot.

If indeed the observed plateau in the inverse slope parameter is due to the softening of the EoS,  the weakening of the collective expansion should also be observed in the longitudinal direction. A prediction for longitudinal pion spectrum can be obtained from Landau's hydrodynamical model~\cite{landau_1, shuryak_1}. In this model the width of the rapidity distribution of pions, $\sigma_{y}(hydro)$, depends on the speed of sound, $c_{s}$,


\begin{equation}
\sigma_{y}^{2}(hydro)=\frac{8}{3}\frac{c_{s}^{2}}{1-c_{s}^{4}}\ln{\left(\frac{\sqrt{s_{NN}}}{2m_{N}}\right)}.
\end{equation}

Using the experimentally measured pion rapidity distributions and Eq. (5) the energy dependence of the speed of sound is obtained in~\cite{bleicher_1}. The distribution shows a well defined minimum confirming the softening of the EoS. 
In the right panel of Fig.~\ref{fig_step} a new version of this plot, namely the ratio of the measured pion rapidity width to the value calculated according to Eq. (5), with $c_{s}^{2} = 1/3$, is shown as a function of $\sqrt{s_{NN}}$. As it is seen from the plot there is a clear minimum indicating the softening of the EoS at the onset energy.

In summary, the experimental status of the structures observed by the NA49 experiment related to the onset of deconfinement was updated in view of the new results from the STAR and ALICE collaborations. While the results of the STAR experiment confirm the measurements of the NA49 collaboration, the new results of ALICE are in agreement with the extrapolations of the measured experimental distributions at lower energies.


\paragraph{Acknowledgments}
The author thanks Marek Gazdzicki and Herbert Str\"obele for useful comments and stimulating discussions. 
The support by the German Research Foundation (grant GA 1480/2.1) is gratefully acknowledged.

\end{document}